# Auditing of AI: Legal, Ethical and Technical Approaches

Jakob Mökander[1,2]

[1] Oxford Internet Institute, University of Oxford, 1 St Giles', Oxford, OX1 3JS, UK

[2] Center for Information Technology Policy, Princeton University, Princeton, NJ 08544, US

Email for correspondence: <jakob.mokander@oii.ox.ac.uk>

**Abstract**

AI auditing is a rapidly growing field of research and practice. This review article, which doubles as an editorial to Digital Society's topical collection on 'Auditing of AI', provides an overview of previous work in the field. Three key points emerge from the review. First, contemporary attempts to audit AI systems have much to learn from how audits have historically been structured and conducted in areas like financial accounting, safety engineering and the social sciences. Second, both policymakers and technology providers have an interest in promoting auditing as an AI governance mechanism. Academic researchers can thus fill an important role by studying the feasibility and effectiveness of different AI auditing procedures. Third, AI auditing is an inherently multidisciplinary undertaking, to which substantial contributions have been made by computer scientists and engineers as well as social scientists, philosophers, legal scholars and industry practitioners. Reflecting this diversity of perspectives, different approaches to AI auditing have different affordances and constraints. Specifically, a distinction can be made between technology-oriented audits, which focus on the properties and capabilities of AI systems, and process-oriented audits, which focus on technology providers' governance structures and quality management systems. The next step in the evolution of auditing as an AI governance mechanism, this article concludes, should be the interlinking of these available – and complementary – approaches into structured and holistic procedures to audit not only how AI systems are designed and used but also how they impact users, societies and the natural environment in applied settings over time.

# 1    Introduction

The prospect of auditing AI systems has recently attracted much attention from researchers, companies and policymakers alike. Following Sandvig et al.'s (2014) article *Auditing Algorithms,* a rich and growing academic literature focuses on how auditing procedures can help identify and mitigate the risks AI systems pose. In parallel, an AI auditing ecosystem is emerging whereby professional services firms like Deloitte (2020), EY (2018), KPMG (2020) and PwC (2020) provide auditing (or 'assurance') services to help clients ensure that the AI systems they design and deploy are ethical, legal and technically robust. This development is not limited to the private sector (Morley et al., 2021). For example, in the *Artificial Intelligence Act* (AIA), the European Commission (2021) sketches the contours of a union-wide auditing ecosystem and mandates 'conformity assessments with the involvement of an independent third party' for specific high-risk AI systems.

But how are we to understand the term 'AI auditing'? In the broadest sense, auditing refers to an independent examination of any entity, conducted with a view to express an opinion thereon (Gupta, 2004). So understood, auditing has a long history of promoting trust and transparency in areas like security and financial accounting (LaBrie & Steinke, 2019). The basic idea is simple: just as the financial transactions of an organisation can be audited for correctness, completeness and legality, so the design and use of AI systems can be audited with respect to not only their technical performance but also their alignment with organisational policies and hard regulations. While this analogy between financial audits and the auditing of AI is useful, it only goes so far. Analogies sometimes constrain our reasoning by uncritically carrying over assumptions from one domain to another (Taddeo, 2016). Hence, a more precise conceptualisation of auditing of AI that makes its functional and operational components explicit is needed.

AI auditing can be defined both with respect to its intended purpose and with respect to its methodological characteristics. Functionally, AI auditing is a governance mechanism that can be wielded by different actors in pursuit of different objectives. For example, AI auditing can be used (i) by regulators to assess whether a specific system meets legal standards, (ii) by technology providers looking to mitigate technology-related risks, and (iii) by other stakeholders wishing to make informed decisions about how they engage with specific companies (Brown et al., 2021). Methodologically, AI auditing is characterised by a structured process whereby an entity's past or present behaviour is assessed for consistency with predefined standards, regulations or norms (Mökander & Floridi, 2021). Figure 1 illustrates how AI auditing is a subset both of AI governance mechanisms (functionally) and auditing procedures (methodologically).

*Figure 1. A schematic overview of how auditing of AI relates to previous work on AI governance*

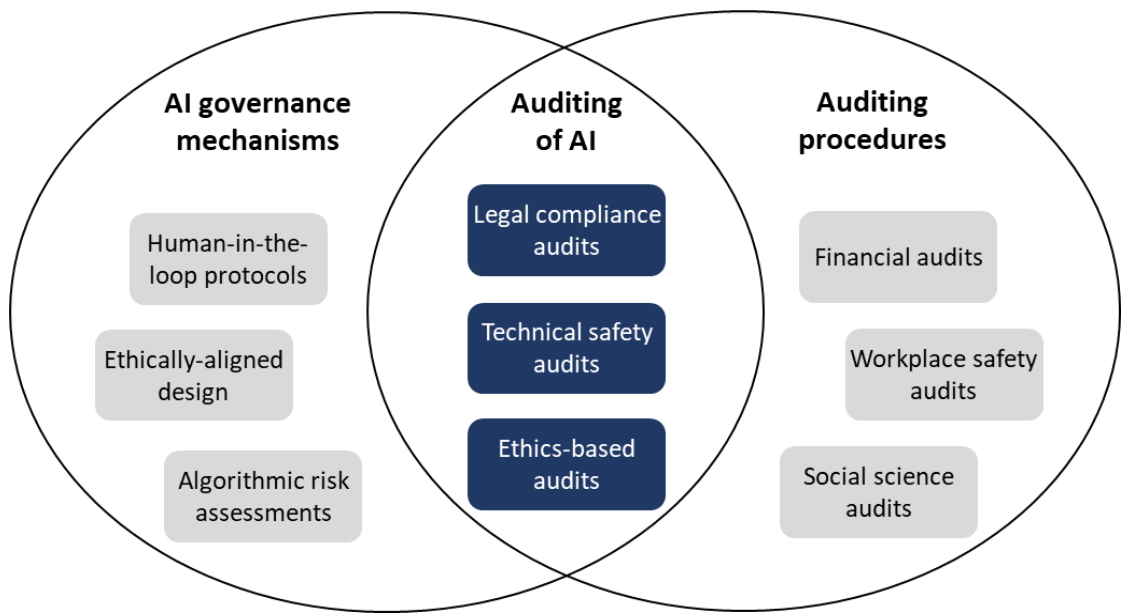

In this review article, I provide an overview of previous work on AI auditing. The literature on AI auditing is at once scarce and rich. It is scarce insofar as AI auditing is a relatively recent phenomenon that few researchers have explicitly addressed – much less studied empirically. In fact, much of the relevant literature has only been published in the last few years (see, e.g., Brown et al., 2021; Metaxa et al., 2021; Mökander et al., 2021; Bandy et al., 2021; Koshiyama et al., 2022; Raji et al., 2020). Still, the literature on AI auditing is rich in the sense that it intersects with almost every aspect of how to govern AI systems – from software development to product testing and verification – and relates to many different academic disciplines, including computer science, social science and legal studies.

This review article serves as an introduction to the journal Digital Society's topical collection on *Auditing of AI: Legal, Ethical and Technical Approaches*. However, rather than summarising the different articles included in the special issue, my aim is to highlight three more general points. First, the theory and practice of AI auditing have only recently begun to mature. While much progress has been made in recent years, I argue that attempts to audit the design and use of AI systems still have much to learn from how audits are structured and conducted in areas like financial accounting, safety engineering and the social sciences.

Second, the contemporary drive towards developing AI auditing procedures results from a confluence of top-down and bottom-up pressures. The top-down pressures consist of forthcoming regulations that reflect governments' needs to manage the ethical and social challenges AI systems pose whilst maintaining incentives for technological innovation. The bottom-up pressures consist of voluntary initiatives that reflect private companies' needs to identify and manage reputational and technology-related risks. In short, both policymakers and technology providers have an interest in promoting auditing as an AI governance mechanism. This, I argue, means that it is left to academic researchers to study how feasible and effective different AI auditing procedures are in practice.

Third, different auditing procedures have different constraints and affordances. Simplified, previous research on AI auditing can be divided into *narrow* and *broad* approaches. The former is technology-oriented and focuses on assessing the outputs of AI for different input data. The latter is process-oriented and focuses on assessing the adequacy of technology providers' quality management systems (QMS). While both strands of research are flourishing, they seldom have dialogue with each other. This is problematic because feasible and effective AI auditing procedures must incorporate elements of both technology- and process-oriented assessments. On the upside, many tools and methods to conduct both types of audits have already been developed. Hence, I argue that the next step in the evolution of auditing as an AI governance mechanism should be to interlink the available tools and methods into structured and independent procedures.

The remainder of this article proceeds as follows. In Section 2, I survey the evolution of auditing as a governance mechanism, discussing how it has been used to promote transparency and accountability in areas like financial accounting and safety engineering. In Section 3, I draw on recent societal developments to show that the need to audit AI systems results from a confluence of top-down and bottom-up pressures. In Section 4, I review previous academic literature in the field of AI auditing. In doing so, I distinguish between narrow and broad conceptions of AI auditing and between legal, ethical and technical approaches to such auditing. In Section 5, I introduce the articles included in this topical collection. Finally, in Section 6, I conclude by showcasing how these articles build on and add to the plurality of auditing procedures that have already been developed to identify and mitigate the risks posed by AI systems.

## 2    The evolution of auditing as a governance mechanism

The promise of auditing as an AI governance mechanism is underpinned by three ideas: that procedural regularity and transparency contribute to good governance (Floridi, 2017); that proactivity in the design of AI systems helps identify risks and prevent harm before it occurs (Kazim & Koshiyama, 2020); and that operational independence contributes to the objectivity and professionalism of the assessment (Raji et al., 2022). However, different researchers and practitioners use the term auditing in different ways. This has caused widespread concern about conceptual confusion in the field (Landers & Behrend, 2022). As Vecchione et al. (2021, p.1) put it:

> *'As [AI] audits have proliferated, the meaning of the term has become ambiguous, making it hard to pin down what audits actually entail and what they aim to deliver.'*

To some extent, such terminological underdetermination is inevitable, given that AI auditing is a fast-moving and multidisciplinary field of research and practice. However, it comes at a cost. Without a shared understanding of what auditing is, let alone widely used standards for how it should be conducted, claims that an AI system has been audited are difficult to verify and may potentially exacerbate rather than mitigate bias and harms (Costanza-Chock et al., 2022). It is therefore useful to take a step back and consider how the term has historically been used in different contexts.

In this section, I briefly review the history of auditing in financial accounting, safety engineering and social science research. The reason for focusing on these specific domains is that, as we shall see, auditing methods and best practices developed in these areas have inspired and informed contemporary attempts to audit AI systems.

## 2.1 Financial audits

The term audit stems etymologically from the Latin *auditus*, meaning 'a hearing'. During Roman times, the term was used with reference to juridical hearings, i.e., official examinations of oral accounts (Lee & Azham, 2008). With time, so-called auditors came to verify written records too. According to Flint (1988), auditing is a means of social control because it monitors conduct and performance to secure or enforce accountability. Auditing is thus a governance mechanism that various parties can employ to exert influence and achieve normative ends. Over time, the objectives and techniques of auditing have developed, reflecting society's changing needs and expectations (Brown, 1962).

The close relationship between auditing and financial accounting is no coincidence. Throughout the Middle Ages, audits were used to verify the honesty of people with fiscal responsibilities (Brown, 1962). However, the rise of financial auditing – as we know it today – stems from shareholders' need to hold professional managers of large industrial corporations accountable. The modern history of auditing thus began in 1844, when the British Parliament passed the Joint Stock Companies Act, which required directors to issue audited financial statements to investors (Smieliauskas & Bewley, 2010). Shortly thereafter, the first public accountancy organisations – which certified independent auditors – were formed in the UK.

Another important transition took place in the 1980s with the rise of risk-based auditing (Turley & Cooper, 2005). Originally, audits were compliance-based in that they sought to verify previously occurring transactions against some pre-established baseline. In contrast, risk-based auditing assessed organisational processes to proactively mitigate risks. Hence, since the 1980s, auditors have not only been expected to enhance the credibility of financial transactions but also provide value-added services like identifying business risks and advising management on how to improve organisational processes (Cosserat, 2004).

In a book titled *The Audit Society*, Power (1997) describes the key aspects of financial auditing procedures, two of which have direct implications for the contemporary discourse on how to audit AI systems. First, Power argues that financial auditing is a 'ritual of verification'. Although auditors examine potential fraud, their primary function is to produce comfort. Similarly, while it may be impossible to mitigate all risks associated with AI systems, systematised audits can promote trust between actors with competing interests through procedural transparency and regularity.

Second, Power argues that the auditor-auditee relationship has multiple layers. On the one hand, auditing presupposes operational independence between auditors and auditees. On the other hand, audits are most effective when the parties collaborate towards a common goal. That tension has created a model called three lines of defence; while management, internal auditors and external auditors should all work to align organisational processes with the interests of different stakeholders, these three actors have complementary roles and responsibilities (IIA, 2009). Recent research suggests that this approach could also help reduce the risks posed by AI systems (Schuett, 2022).

To summarise, financial auditing and accounting has grown into one of the world's largest industries, with an estimated market size of over $110bn (Grand View Research, 2017). Consequently, the industry is highly professionalised. Many organisations with roots in that industry have utilised their know-how and strong market positions to expand horizontally by offering other auditing services. As a case in point, the Institute of Internal Auditors (IIA, 2018) has recently developed a framework for how to audit AI systems. Similarly, professional services firms that have historically focused on financial audits have now started to offer clients AI auditing services too.

## 2.2 Safety audits

Although the modern history of auditing started with financial audits, safety audits represent an equally well-established area of theory and practice. While the former seeks to manage financial risks, the latter aims to highlight health and safety hazards and assess the effectiveness of the mechanisms in place to address them (Allford & Carson, 2015). Examples include workplace safety audits (Gay & New, 1999), food safety audits (Dillon & Griffith, 2001) and operation safety audits in the aviation industry (Klinect et al., 2003). The history of safety audits stretches back to the Industrial Revolution in 19th-century Britain. At that time, the conditions for workers were poor and the risk of injury or death following workplace accidents was high (Frey, 2019). With time, however, workers formed unions demanding better conditions. One of the mechanisms institutionalised to hold employers accountable was workplace safety auditing. Allford and Carson (2015, p.1) defined the practice thus:

> *'Safety audits check that what the business does in reality matches up to both what it says it does [according to its own policies] and what it [legally] should do to continuously ensure that major accident risks are reduced as much as possible.'*

Safety audits hold valuable lessons for how to design feasible and effective auditing procedures. First, safety auditors rely on a plurality of tools (e.g., checklists) and methods (e.g., interviews) to assess the adequacy of organisational safety management systems (Kuusisto, 2001). The lesson that different auditing procedures must not be seen as mutually exclusive but rather complementary holds true for AI auditing as well. Second, no audit is stronger than the institutions backing it. Safety audits are conducted by independent auditors, who belong to or are certified by NGOs like the British Safety Council or government bodies like the US's Occupational Safety and Health Administration. An equally rigorous institutional ecosystem to conduct and enforce AI audits has yet to emerge (ICO, 2020). Finally, safety audits highlight the interdependence between technical and social systems. Most accidents involving engineered systems do not stem from the failure of technical components but from requirement flaws or handling errors (Leveson, 2011). The main objective of safety audits is thus to assess and improve organisations' safety cultures. This implies that AI audits must also consider the culture within organisations designing or deploying such systems.

Despite their merits, safety audits have limitations as a governance mechanism. For example, the history of food safety demonstrates that audits can reduce but never eliminate the risk of incidents occurring (Powell et al., 2013). Moreover, safety auditing may become a box-ticking exercise, which not only wastes resources but can also create a false sense of security that increases the risk of adverse events (Allford & Carson, 2015). Finally, because safety auditors rely on auditees' active cooperation, they often struggle to access the required evidence. This final limitation is likely to be a concern for AI auditors too since their access tends to be limited by intellectual property rights and privacy legislation.

While financial and safety audits differ in substance, they share both procedures and functions. In both cases, auditors seek to verify auditees' claims with the dual aim of reducing risks and providing a basis for holding management accountable. However, as the history of social science audit studies shows, the term auditing has been used rather differently in other contexts.

### *2.3 Audit studies in the social sciences*

In the social sciences, the term 'audit study' refers to a research method, specifically a type of field experiment, which is used to examine individuals' behaviour or the dynamics of social processes (Gaddis, 2018). Field experiments attempt to mimic natural science experiments by implementing a randomised research design in a real-world setting (Baldassarri & Abascal, 2017). The advantage of field experiments – compared to surveys or interviews – is that they allow researchers to study people and groups in their natural environment. Gaddis defined an audit study as follows:

> *'Audit studies [in the social sciences] generally refer to a specific type of field experiment in which a researcher randomizes one or more characteristics about individuals and sends these individuals out into the field to test the effect of those characteristics on some outcome.'* Gaddis (2018, p.5)

Audit studies have been employed by social scientists since the 1950s, often to examine difficult-to-detect behaviours, such as racial and gender discrimination. For example, Bertrand and Mullainathan (2004) investigated racial discrimination in hiring across a wide range of sectors by designing an audit study in which they drafted and submitted fictitious résumés in response to job postings. They varied white-sounding and black-sounding names on similar résumés and measured the responses to those applications. Résumés with white-sounding names were 50% more likely to get callbacks from interviewers than those with black-sounding names.

Many similar social science audit studies have been conducted. Although sharing a basic methodology, these studies vary in two dimensions. The first is the domain being studied. Beyond recruitment, audit studies have been conducted in areas like healthcare (Kugelmass, 2016) and social housing (Ahmed & Hammarstedt, 2008). The second dimension is the choice of independent variable, i.e., the characteristic being manipulated by the researchers. In addition to race, the design of audit studies has included manipulation of gender (Neumark et al., 1996), age (Farber et al., 2017) and religion (Pierné, 2013), just to mention a few examples.

The social science audit study is a suitable methodology for gathering information about discrimination caused by AI systems too. In fact, this is already happening. Several examples of algorithmic audits are of this kind, including Boulamwini and Gebru's (2018) audit study, which demonstrated that AI systems used to classify images of people according to gender were significantly more accurate when applied to lighter-skinned males than darker-skinned females. In Section 4, I will return to the literature on social science audits focusing specifically on AI systems. Here, I wish to make a further distinction that aids understanding of the different strands of auditing research.

There are many ways to conduct social science research. For example, there is a long-standing methodological tension between explanation-oriented research seeking to gather empirical evidence on social phenomena and activist research striving to advance a specific normative agenda or change the material conditions of the people and places being studied (Hale, 2017). Both approaches have merits and – as the philosophy of social science has shown – are not mutually exclusive but overlap in practice (Cartwright & Montuschi, 2014). However, how researchers relate to their object of study matters, and the field of auditing is no different.

Historically, audit studies in the social sciences have been associated with so-called activist research. Cancian (1993) defines activist research as research that aims to promote changes that equalise the distribution of resources by exposing inequalities. Audit studies conducted by activist researchers tend to be adversarial in nature, seeking to highlight injustices in ways that spark reactions. In contrast, audits conducted by professional service providers in industry settings aim to produce comfort (Power, 1997). There are thus deep tensions in the motivations different practitioners and researchers have for conducting audits. As the next section will show, these tensions also persist in the literature on AI auditing too.

## 3  The need to audit AI systems – a confluence of top-down and bottom-up pressures

Auditing procedures are institutionalised in response to the perceived needs of individuals and groups who seek information or reassurance about the conduct or performance of others in which they have legitimate interests (Flint, 1988). In Section 2, that point was illustrated by describing how financial audits emerged in response to investors' needs and how safety audits were institutionalised in response to social and political pressures to improve working conditions. In the introduction to this article, I stressed that AI auditing is not just a theoretical possibility but already a widespread practice. That sparks two questions: to which perceived needs do these auditing procedures respond? And which stakeholders are seeking information or reassurance through auditing of AI systems?

In this section, I argue that the need to auditing AI systems results from a confluence of top-down and bottom-up pressures. The former includes the regulatory mandates and normative expectations placed on technology providers by external stakeholders like policymakers and advocacy groups. The latter includes voluntary measures taken by technology providers to stay competitive in their industries, including continuous improvements in software development and testing procedures. Figure 2 illustrates how this confluence of top-down and bottom-up pressures results in a growing need to audit AI systems. In what follows, I will discuss these pressures in turn.

*Figure 2. The need to audit AI systems is underpinned by both top-down and bottom-up pressures*

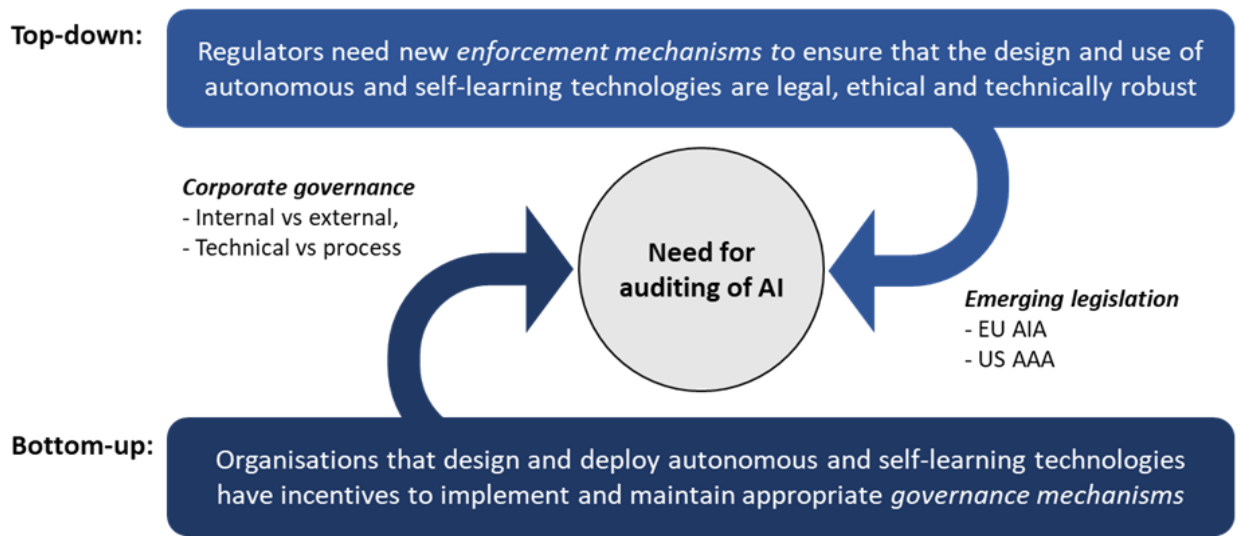

### 3.1 Auditing as a mechanism to implement legislation

A major driver behind the proliferation and implementation of AI auditing procedures is forthcoming government regulations. To appreciate the force behind this top-down pressure, it is useful to take a step back. AI systems have great potential to contribute to both economic growth and human well-being. By drawing inferences from the growing availability of (big) data, AI systems can improve the speed and accuracy of information processing and contribute to the development of new innovative solutions (Taddeo & Floridi, 2018). However, the ethical, social and legal challenges AI systems pose are equally evident. AI systems may not only cause harm related to bias, discrimination and privacy violations but also enable human wrongdoing and undermine self-determination (Tsamados et al., 2021). Policymakers are thus faced with the challenge of balancing the prevention of harm against providing incentives for innovation.

Consider recent developments in the field of large language models (LLMs) as an example. The release of ChatGPT has drawn public attention to the capacity of LLMs – such as OpenAI's GPT-3 (Brown et al., 2020) and Google's LaMDA (Thoppilan et al., 2022) – to generate human-like text based on the input provided to them. While such texts are not always semantically meaningful, they can still be used for tasks like text summarisation and translation (Floridi & Chiriatti, 2020). Yet there has been a strong backlash against how LLMs are designed and used. Some researchers have shown that LLMs can produce unethical language, including racist and sexist comments (Kirk et al., 2021). Others have proved that LLMs' answers often contain factual errors (Evans et al., 2021).

The seriousness of these limitations is exacerbated by the fact that open-source business models allow LLMs to be used for tasks they were not originally designed to perform (Bommasani et al., 2021). For instance, in January 2023, a Columbian judge used ChatGPT to transcribe his interactions with witnesses, material that he later used to justify his verdict (Parikh et al., 2023). This and other similar examples have understandably sparked widespread public outcry (Kak & West, 2023). Of course, it is important not to be carried away by the latest technological innovation or regulatory trends. Still, the case of LLMs illustrates a more general point, namely, that policymakers are facing increasing pressure to regulate the design and use of AI systems (Smuha, 2021).

In many jurisdictions, this has meant drafting new legislation. Published in April 2021, the European AIA was the first comprehensive regulatory framework for AI systems proposed by any major global economy. However, already before that governments had proposed more targeted legislation. For example, the Government of Canada (2019) has published a *Directive on Automated Decision-Making*, and the Government of Singapore (2020) has published guidelines on how to design and use AI systems responsibly. A similar bill labelled the *Algorithmic Accountability Act of 2022* (AAA) is currently being considered by the US Congress (Mökander & Floridi, 2022a). These draft regulations differ in scope and substance. However, they all stipulate rules and requirements that organisations designing or deploying AI systems must follow. In some cases, the focus is on substantive requirements. For example, AI systems used as components in medical devices must meet specific performance standards in both the EU (Niemiec, 2022) and the US (FDA, 2021). In most cases, however, the focus is on process-based rules (Veale & Borgesius, 2022).

Whatever form they take, regulations must be linked to effective governance mechanisms to be implemented and enforced (Baldwin & Cave, 1999). For example, the AIA threatens technology providers that fail to comply with its requirements with hefty fines (European Commission, 2021). However, to determine compliance, one must first consider what mechanisms are available to establish what a provider is doing. This is where auditing comes in. As financial transactions can be audited for correctness, completeness and legality, so the design and use of AI systems can be audited for technical robustness and legal compliance.

This development is already well underway. The EU AIA, for instance, mandates that high-risk AI systems undergo conformity assessments before deployment. By demanding that these assessments are conducted in a structured manner by independent third parties that have been accredited by national authorities, the European Commission is sketching an EU-wide auditing ecosystem in all but name (Mökander et al., 2022). In addition to the AIA, the EU has recently published guidance on how to audit the quality of datasets used in algorithmic decision-making (EPRS, 2022). Similarly, the UK Information Commissioner's Office (ICO, 2020) has issued guidance on how to audit AI systems. However, the most mature government regulation is currently found in the US. In 2021, New York City enacted the *AI Audit Law* (NYC Local Law 144), requiring that AI systems used to inform employment-related decisions are made subject to independent audits:

> *'New York City's law will restrict employers from using AI systems in hiring and promotion decisions unless it has been the subject of a bias audit by an independent auditor no more than one year prior to use.'* (Gibson Dunn, 2023, p.1)

As these examples illustrate, audits can be used by regulators seeking to assess whether an AI system is legally compliant. Like financial audits, AI audits thus respond to one actor's perceived need to gather information about another's conduct. However, there is a major difference between financial audits and legally mandated AI systems audits. Investors exert pressure on managers motivated by the need to manage financial risk. In contrast, policymakers exert pressure on technology providers (in part) to maintain political legitimacy. As noted by Peter (2010), a government's legitimacy hinges partially on its success in solving social and economic problems. As ever more critical tasks become automated, policymakers' political legitimacy will increasingly depend on their abilities to manage the ethical and social challenges AI systems pose. Consequently, the top-down pressure to institutionalise procedures to audit AI systems is likely to continue accumulating. However, as we shall see, such pressure is not the only driver behind the emergence of a new AI auditing industry.

### 3.2 The role of AI auditing in corporate governance

Private companies play a major role in designing and deploying AI systems (Cihon et al., 2021). Therefore, their design choices have direct and far-reaching implications for important issues, including social justice, economic growth and public safety (Baum, 2017). However, the dominance of private sector actors holds true not only for the development of commercial applications but also for basic research on the computational techniques that underpin the capabilities of AI systems. For example, in 2018, private companies and labs published over 50% more research papers on ML than academics in the US (Perrault et al., 2019). Hence, the policies and governance mechanisms private companies employ to guide their design and use of AI systems are of profound societal importance.

In the previous section, I showed that policymakers have reasons for mandating audits of AI systems. However, previous research suggests that technology providers too have strong incentives to subject the AI systems they design and deploy to independent audits (Falco et al., 2021; Raji et al., 2020). To understand those incentives, it is useful to first consider the function of corporate AI governance, which Mäntymäki et al. define as follows:

> *'AI governance is a system of rules, practices, processes, and technological tools that are employed to ensure that an organization's use of AI systems aligns with the organization's strategies, objectives, and values.'* (Mäntymäki et al., 2022, p.2)

As this definition suggests, corporate governance seeks to ensure that the conduct of an organisation aligns with its stated objectives (OECD, 2015). However, the environment in which corporate governance takes place is inherently dynamic (Arjoon, 2005). As Schumpeter (1942) argued, private companies face constant pressures to innovate and improve their products. Technology providers have thus developed mechanisms to ensure that their products and services meet predefined quality standards and respond to consumers' needs. Since both the underlying technologies and consumer needs keep changing, the mechanisms employed to govern organisational processes must also be continuously revised.

This brief detour into the function of corporate governance has direct implications for why technology providers voluntarily subject themselves and their AI systems to audits. As noted by Russell et al. (2015), questions concerning corporate AI systems governance are of two kinds: (i) did we build the system right? And (ii) did we build the right system? The former is a technical question; the latter is a normative one. Audits can provide answers to both kinds of question, as two real-world examples illustrate.

O'Neil (2016) told the story of a woman who, despite a competitive CV, could not get a job due to an error in the algorithmic vetting system used by many recruiters. It was eventually revealed that an alleged criminal offence in her file originated from a data-scraping program, which had conflated her and someone with the same name and postcode. This shows the dangers of negligent design, irresponsible data management, and questionable deployment of AI systems. It is important to note, however, that in this case the data controller, employer and job seeker would all have benefited from a 'correct' classification. This type of poor-quality outcome constitutes a technical problem that developers, at least in theory, can address. To do so, developers need to be both made aware of the limitations of the AI systems they design and incentivised to act on that information.

This is where auditing comes in. By assessing the capabilities and limitations of AI systems prior to deployment, auditing helps technology providers identify and mitigate risks before harm occurs (Wilson et al., 2021). Further, by providing a basis on which technology providers can be held accountable, audits incentivise investments in adequate risk management (Shen et al., 2021). In fact, one of the main reasons organisations subject themselves to independent audits is to assess and improve their software development processes and QMS (Vlok, 2003). After all, it is often cheaper to address vulnerabilities early in software development processes. Dawson et al. (2010) estimated that it can cost up to 15 times more to fix a bug in an AI system when it is found during the testing phase rather than the deployment phase.

In other cases, however, public outcry has been directed not against the technical failures of AI systems but against the purposes for and ways in which they were built in the first place (Keyes et al., 2019). In 2020, Clearview AI – a facial recognition company – faced backlash after investigations revealed that it had scraped billions of images from social media platforms without users' consent to assemble its training dataset (Hill, 2020). Clearview AI suffered significant reputational damage (Smith & Miller, 2022) and faced legal actions culminating in a settlement banning it from selling its technologies to private companies in the US (Robertson, 2022). While it remains unclear whether Clearview AI violated the law, it evidently violated customers' and citizens' normative expectations.

This brings us to the second point: audits focusing on not only technical but also ethical aspects of AI systems help technology providers manage financial and reputational risks (EPRS, 2019). Proactive communication of audit findings may help companies gain competitive advantages: just as organisations seek to show consumers that their products are healthy through detailed nutritional labels (Holland et al., 2018), the documentation of steps taken to ensure that AI systems are ethical can play a positive role in both marketing and public relations. Specifically, previous research suggests that structured and independent audits of AI systems can help organisations improve on several business metrics like regulatory preparedness, data security, talent acquisition, reputational management and process optimisation (EIU, 2020; Schonander, 2019).

In light of these bottom-up pressures, it is unsurprising that many technology providers have already voluntarily implemented procedures to audit their AI systems for alignment with different sets of ethics principles. Yet this development also calls for caution. Sloane (2021) argued that audits commissioned by technology providers are insufficiently independent, and Bandy (2021) pointed out that, in the absence of agreed standards, technology providers' claims that their AI systems have been audited are hard to verify. These objections should be taken seriously. However, this section has not sought to assess the merits of AI systems auditing as a governance mechanism but only to highlight that both policymakers and technology providers have an interest in developing and promoting procedures to audit such systems. The study of how feasible and effective these auditing procedures are in practice is an exercise best left to academic researchers.

## 4  Auditing of AI's multidisciplinary foundations

In this section, I review what I refer to as the AI systems auditing literature. What unites all works in this body of literature is that they concern procedures to audit AI systems for consistency with relevant specifications, regulations or ethics principles. However, before proceeding further, it is useful to revisit and expand the definition of AI auditing provided in the introduction.

### *4.1  The AI auditing literature*

To recap, AI auditing can be defined both functionally and methodologically. Functionally, AI auditing is a governance mechanism that can be wielded by different actors in society in pursuit of different goals and objectives. For example, it can be used by regulators to assess whether a specific AI system meets legal standards, by technology providers to mitigate technology-related risks, or by other stakeholders to make informed decisions about how they engage with specific companies (Brown et al., 2021). Methodologically, auditing of AI systems is characterised by a structured process whereby an entity's past or present behaviour is assessed for consistency with predefined standards, regulations or norms.

Four aspects of this definition of AI auditing require further clarification. First, the subject of the audit can be either a person, an organisation, a technical system or any combination thereof (Mökander & Axente, 2021). Second, different auditing procedures follow different logic. *Functionality audits* focus on the rationale behind decisions; *code audits* entail reviewing the source code of an AI system; and *impact audits* investigate the types, severity and prevalence of effects of an AI system's output (Mittelstadt, 2016). Importantly, these distinct approaches are not mutually exclusive but rather crucially complementary. Third, whether conducted by an *external* third party or an *internal* audit function, auditing requires operational independence between the auditor and the auditee (Power, 1997). Finally, auditing requires a predefined baseline to serve as a basis for evaluation (ICO, 2020). However, the nature of this baseline can vary between hard regulations, organisational values and policies, or technical standards and benchmarks.

Previous work on AI systems auditing constitutes a heterogeneous and multidisciplinary body of literature. It is heterogeneous in that it encompasses contributions from a diverse range of actors employing different methods and facing competing incentives. The AI systems auditing literature includes academic articles and books (Berghout et al., 2023), auditing tools and procedures developed by private companies (Babl AI, 2023; ORCAA, 2020), standards published by industry associations and professional standard-setting bodies (IEEE, 2019; ISO, 2022; NIST, 2022; VDE, 2022), and draft legislation and guidance documents issued by policymakers (EPRS, 2022; European Commission, 2021; ICO, 2020), to mention just a few examples.

The AI systems auditing literature is also multidisciplinary in that it harbours contributions from many academic disciplines, including computer science (Adler et al., 2018; Kearns et al., 2018), systems engineering (Dennis et al., 2016; Leveson, 2011), law (Laux et al., 2021; Selbst, 2021), media and communication studies (Bandy & Diakopoulos, 2019; Sandvig et al., 2014), social science (Metaxa et al., 2021; Vecchione et al., 2021), philosophy (Boddington, 2017; Dafoe, 2017) and organisational studies (Guszcza et al., 2018).

Such a diverse body of literature can be sliced and diced in many ways. In what follows, I provide an overview of the AI systems literature in three steps. First, I distinguish between narrow and broad conceptions of auditing. Second, I distinguish between technical, legal and ethical approaches to AI systems auditing. Finally, I distinguish between strands of research that (i) propose, (ii), develop, (iii) employ or (iv) critique AI systems auditing procedures.

### 4.2 Narrow vs broad conceptions of auditing of AI systems

To start with, it is useful to distinguish between narrow and broad conceptions of AI auditing. The former is impact-oriented, focusing on probing and assessing the output of AI systems for different input data. The latter is process-oriented, focusing on assessing the adequacy of the software development processes and QMS technology providers employ.

In their book *Auditing Algorithms: Understanding Algorithmic Systems from the Outside In*, Metaxa et al. provided an example of a narrow definition of auditing:

> *'[an algorithm audit is] a method of repeatedly and systematically querying an algorithm with inputs and observing the corresponding outputs in order to draw inferences to its opaque inner workings.'* (Metaxa et al., 2021, p.18)

Narrow conceptions of auditing are well suited to gathering evidence about unlawful discrimination and tend to be underpinned by experimental designs. For example, in an article titled *Algorithm Auditing at Large-Scale: Insights from Search Engine Audits*, Ulloa et al. (2019) designed virtual agents to perform systematic experiments simulating human interactions with search engines. The authors demonstrated that such an audit design can be employed to monitor an AI system's output over time and flag potential ethical concerns such as disparate treatment.

In contrast, broad conceptions of auditing focus not so much on the properties of AI systems as the governance structures of the organisations that design and deploy them. This practice has deep roots in conventional IT audits Zinda (2021) and technology risk management procedures (Senft & Gallegos, 2009). Jager and Westhoek describe the role of such an auditor:

> *'It is not just about checking the algorithm itself and the management measures surrounding it, but also paying attention to the data used, the methods used in the development and the optimization of the algorithm. These aspects of management, process, and content should also be part of the assessment framework and thus the audit approach.'* (Jager & Westhoek, 2003, p.145)

Broad conceptions of auditing are useful since they allow researchers not only to detect the illegal, erroneous, or unethical behaviours of AI systems but also to investigate the sources of such behaviours. For example, discriminatory behaviour of AI systems may be caused by incomplete or unrepresentative training datasets (Gehman et al., 2020) or inadequate AI systems testing and validation procedures (Myllyaho et al., 2021). For this reason, researchers like Koshiyama et al. (2022) have proposed procedures for auditing the entire process whereby AI systems are designed and deployed. Typically, this entails assessing the governance structures technology providers have in place to train their staff, assemble training datasets, evaluate the limitations of AI systems prior to deployment, and monitor the behaviour of AI systems over their entire lifetime.

Both narrow and broad conceptions of auditing have generated flourishing strands of research. Some researchers have leveraged narrow conceptions of auditing to test for bias and discrimination in online ad delivery (Ali et al., 2019; Sweeney, 2013) and autocomplete algorithms (Robertson et al., 2018), for fairness in image classification systems (Morina et al., 2019), for accuracy in news curation systems (Bandy & Diakopoulos, 2019), for completeness in datasets (Coston et al., 2021; Sookhak et al., 2014), and for data privacy, e.g., how easy it is to reconstruct training data from AI systems (Kolhar et al., 2017; Narula et al., 2018).

Other researchers have leveraged broad conceptions of auditing to study how AI systems are designed and the adequacy of technology providers' governance mechanisms. Ugwudike (2021) studied how AI systems used for predictive policing are designed and deployed; Jager and Westhoek (2023) studied technology providers' mechanisms for testing image recognition algorithms; Mahajan et al. (2020) provided a framework for how auditors and vendors can collaborate to validate AI systems used in radiology; and Dash et al. (2019) demonstrated how audits of recommender systems can provide insights into how these systems affect users and societies over time.

This discussion has two key takeaways. First, narrow and broad conceptions of auditing have different affordances. The former allows researchers to audit the behaviour of AI systems without approval from, or the cooperation of, technology providers (Adler et al., 2018; Lee, 2021; Lurie & Mustafaraj, 2019). The latter enables researchers to study the real-world effects different auditing procedures have on how AI systems are designed and deployed (Ayling & Chapman, 2021; Fitzgerald et al., 2013; Stoel et al., 2012). Second, there is no contradiction between the two concepts. In fact, they are both compatible and mutually reinforcing. Specifically, narrow testing of AI systems based on input-output relationships can (and should) be integrated into broader auditing procedures.

*4.3 Technical, legal and ethics-based approaches*

In addition to having different methodological conceptions of what auditing is, researchers also differ in what they are auditing AI systems *for*. Per definition, auditing requires a predefined baseline against which the audit's subject can be evaluated (ICO, 2020). However, depending on the audit's purpose, this baseline can consist either of technical specifications, legal requirements or voluntary ethics principles. Consequently, contributions to the AI systems auditing literature can be categorised into technical, legal and ethical approaches.

The term technical approaches refers to auditing procedures designed to quantify and assess the technical properties of AI systems, including accuracy, robustness and safety. These build on tools and methods with proven track records in systems engineering and computer science, including model evaluation (Parker, 2020) and system verification (Luckcuck et al., 2019; Thudi et al., 2021). Within the realm of technical approaches, a distinction is often made between ex-ante and ex-post audits (Etzioni & Etzioni, 2016). The former evaluates an AI system prior to its market deployment, the latter monitors its performance over time as it interacts with new input data in applied settings.

The idea of auditing software dates back several decades (Hansen & Messier, 1986; Weiss, 1980). Still, the academic literature in this field has grown rapidly in recent years. Some research groups have developed open-source toolkits allowing technology providers to test and evaluate the performance of AI systems on different tasks and datasets (Cabrera et al., 2019; Saleiro et al., 2018). Others have developed auditing procedures for more targeted purposes, e.g., to test the accuracy of personality prediction in AI systems used for recruitment (Rhea et al., 2022), evaluating the capabilities of language models (Goel et al., 2021; Mökander et al., 2023), providing explanations for black-box AI systems (Pedreschi et al., 2018), and conducting audits of clinical decision support systems (Panigutti et al., 2021). Again, what links all these procedures is that they audit AI systems against predefined technical, functionality and reliability standards.

In contrast, the term legal approaches refers to auditing procedures that assess whether the design and use of AI systems comply with relevant regulations. Such procedures rely on different legal provisions, including those stipulated in data privacy regulations like the European Parliament's (2016) General Data Protection Regulation (GDPR), discrimination laws like the US's 1964 Civil Rights Act or Equal Credit Opportunity Act of 1974 (Barocas & Selbst, 2016), sector-specific certification mandates, as is the case for medical device software (FDA, 2021), or general transparency obligations, such as those found in the AIA (European Commission, 2021). Legal scholars have debated about when and how the above-listed regulations apply to AI systems (Durante & Floridi, 2021; Edwards & Veale, 2018; Pentland, 2020; Wachter et al., 2017).

A wide range of procedures to audit AI systems for legal compliance have already been proposed and, in some cases, implemented (Merrer et al., 2022). For instance, Mikians et al. (2012) developed a procedure to audit AI systems for unlawful price discrimination based on protected attributes. Similarly, Silva et al. (2020) audited Facebook's ad delivery algorithm, finding that it violated political advertising laws.

Finally, the term ethics-based approaches refers to auditing procedures for which voluntary ethics principles serve as the normative baseline. Ethics-based auditing can be either collaborative or adversarial. In the former case, audits are conducted in collaboration with technology providers to assess whether their AI systems adhere to predefined ethics principles (Berghout et al., 2023; Raji et al., 2020). In the latter case, independent actors conduct audits to assess an AI system without access to its source code (Sandvig et al., 2014). Collaborative audits aim to provide assurance, adversarial audits to expose harms. In both cases, however, ethics-based auditing concerns what ought to be done over and above compliance with existing regulations.

In ethics-based procedures, AI systems are audited against either a technology provider's organisational values or ethics principles proposed by institutions like the IEEE (2019), OECD (2019) and the AI HLEG (2019). While these guidance documents vary in language (Jobin et al., 2019), they converge on a limited set of principles (Floridi & Cowls, 2019). Reflecting that convergence, previous research has developed procedures to audit AI systems for transparency and explainability (Cobbe et al., 2021; Mittelstadt, 2016), bias and fairness (Bartley et al., 2021; Morina et al., 2019) and accountability (Busuioc, 2021; Metcalf et al., 2021). Many private companies have already subjected themselves to ethics-based audits. Take AstraZeneca as an example. In 2021, the biopharmaceutical company contracted an independent third-party auditor to assess whether the company's use of AI systems to improve drug development processes aligned with its publicly stated AI ethics principles (Mökander & Floridi, 2022b).

In practice, the boundaries between technical, legal and ethics-based audits are often blurry. To demonstrate legal compliance, auditors typically rely on technical methods for gathering evidence about the properties and impact AI systems have (Kim, 2017). Similarly, technical robustness and legal compliance are often prerequisites for considering an AI system ethical (Keyes et al., 2019). The three audit types are thus best viewed as a continuum of complementary approaches with different focal points. That said, the distinction between technical, legal and ethical approaches is useful for two reasons. First, it mirrors the vocabulary adopted by policymakers. For example, AI HLEG (2019) stipulated that AI systems should be lawful, ethical, and technically robust. Adopting this well-established vocabulary facilitates communication with my target audiences. Second, it helps distinguish different types of audits that serve different purposes.

### 4.4  *Who audits the auditors?*

Contributions to the academic literature on AI systems auditing relate to the object of study in different ways. For example, distinctions can be made between contributions that (i) provide theoretical justifications for why audits are needed, (ii) develop procedures, tools or methods to audit AI systems, (iii) employ available auditing procedures, tools or methods, and (iv) study the effectiveness and feasibility of auditing AI systems as a governance mechanism. In what follows, I briefly review these different research strands.

To start with, there is a significant body of literature calling for AI systems to be audited (Diakopoulos, 2015; Sandu et al., 2022; Sandvig et al., 2014). These contributions stress the social, ethical and legal risks AI systems pose and how audits can help identify and manage those risks. For example, research has suggested that auditing contributes to good governance through procedural regularity and transparency (Floridi, 2017b; Larsson, 2020; Loi et al., 2020) and prevents harm by ensuring proactivity in the design of AI systems (Kazim & Koshiyama, 2020). Such contributions are often commentary or viewpoint articles (Falco et al., 2021; Guszcza et al., 2018; Kassir et al., 2022). The main argument advanced by this literature is that structured and independent audits constitute a pragmatic approach to managing the governance challenges of AI systems.

Responding to these calls, other researchers have developed tangible AI systems auditing procedures and tools. Such contributions can be divided into two broad categories. First, high-level procedures – often proposed by scholars from organisation studies or systems engineering – that outline the steps audits should include, what activities these entail, and the roles and responsibilities of different stakeholders (Floridi et al., 2022; Zicari et al., 2021). Second, tools that can be employed by auditors for specific tasks, including detecting bias in AI systems (Saleiro et al., 2018; Sokol et al., 2022), documenting how AI systems are designed (Gebru et al., 2021; Mitchell et al., 2019), and simulating or monitoring their behaviour in real-world settings (Akpinar et al., 2022). These tools are typically developed by computer scientists or social scientists.

Yet other researchers employ existing auditing procedures and tools to conduct empirical studies (Aragona, 2022), including qualitative studies that assess how AI systems are designed (Christin, 2020; Marda & Narayan, 2021; Seaver, 2017) and quantitative audit studies that measure the properties of AI systems or their impact on users and societies (Abebe et al., 2019; Speicher et al., 2018). Contributions to this literature have been made by researchers from different fields. For example, labour economist Songül Tolan (2019) audited AI systems used by courts to predict criminal recidivism and found they discriminate against male defendants and people of specific nationalities. A team of computer scientists led by Alicia DeVos et al. (2022) conducted user-centric audits to study AI systems, concluding that users were able to identify harmful behaviours that formal testing processes had not detected.

Finally, a small but growing community of researchers are interested in how feasible and effective auditing is as an AI system governance mechanism (Costanza-Chock et al., 2022; Landers & Behrend, 2022). So far, such research has been dominated by theoretical critiques. For example, Sloane (2021) argued that current auditing procedures are toothless and may even be counterproductive insofar as they legitimise the deployment of potentially harmful AI systems. To avoid that trap, Sloane suggested that standards for how to audit AI systems are urgently needed. Similarly, Engler (2021) argued that independent auditors struggle to hold technology providers accountable because – in the absence of sector-specific legislation – they can simply refuse access to their data and models. These important objections call for further inquiry. As of now, however, claims about the limitations of AI systems auditing as a governance mechanism have yet to be substantiated by empirical research (just as claims about its affordances).

## 5  In this topical collection

As this review article has aimed to show, AI auditing is a rapidly growing field of research and practice. However, well-established standards for AI auditing have yet to emerge. Further, there remains a large discrepancy between the attention that AI auditing has attracted, on the one hand, and the lack of empirically grounded academic research concerning the effectiveness and feasibility of different auditing procedures, on the other. To help bridge these gaps, Digital Society has published a topical collection titled *Auditing of AI: Legal, Ethical, and Technical Approaches*. The six articles included in the collection speak best for themselves. Hence, the aim of this section is not to summarise each article but only to highlight their contributions in relation to previous research.

As stressed throughout this article, there is a gap between principles and practice in AI auditing. Three contributions to the topical collection address that gap by documenting and reflecting on the challenges and best practices associated with designing and conducting AI audits.

In *Algorithmic Bias and Risk Assessments: Lessons from Practice*, Hasan et al. (2022) help bridge that gap by documenting and reflecting on the challenges auditors and industry practitioners face when designing and conducting AI audits. The article differs from previous research insofar as its findings are based not on reasoning from first principles but on the authors' own experience from advising and conducting AI audits for clients across different industries over the last four years. The article highlights the importance of designing audits in ways that situate AI systems in their proper context, i.e., as components in larger socio-technical systems. Specifically, Hasan et al. describe how 'broad' ethical risk assessment and more 'narrow' technical algorithmic bias assessment depend on and complement each other. The article thus points to an important avenue for future research: how to combine available tools and methods into holistic and structured auditing procedures.

In *Achieving a Data-Driven Risk Assessment Methodology for Ethical AI*, Felländer et al. (2022) outline a cross-sectoral approach for ethically assessing and guiding the development of AI systems. Specifically, the authors propose a data-driven risk assessment methodology for ethical AI (DRESS-eAI). Based on the ISO 31000:2009 risk management process, DRESS-eAI spans six phases: (i) problem definition, (ii) risk scanning, (iii) risk assessment, (iv) risk mitigation, (v) stakeholder engagement, and (vi) AI sustainability reporting. While similar frameworks have been proposed in the past, Felländer et al.'s main contribution is to provide detailed guidance on how to implement DRESS-eAI and what activities each phase entails. Hence, the article is not only relevant to academics and auditors but also to organisations seeking pragmatic guidance on how to ensure and demonstrate that the AI systems they design or deploy adhere to predefined principles.

A further AI auditing procedure, Z-Inspection, is presented and discussed by Vetter et al. (2023) in *Lessons Learned from Assessing Trustworthy AI in Practice.* Z-Inspection is a holistic and dynamic framework to evaluate the trustworthiness of AI systems at different stages of their lifecycle.

The procedure focuses on identifying and deliberating on ethical issues and tensions through the analysis of socio-technical scenarios. The authors illustrate how Z-Inspection works through real-world examples of its application to assess AI systems used in the healthcare sector and for environmental monitoring purposes. A key feature of Z-Inspection is that it allows for the inclusion of various experts from different backgrounds and provides a structured way for them to find an agreement through a require-based framework. The downside of such a procedure is that it is time-consuming and requires subject matter expertise. The upside is that it allows developers and users of AI systems to address specific ethical issues in applied settings.

The articles hitherto discussed all focus on the practical implementation of AI auditing. However, other contributions are conceptual in nature. In *Continuous Auditing of Artificial Intelligence: a Conceptualization and Assessment of Tools and Frameworks,* Minkkinen et al. (2022) revisit the concept of continuous auditing – as conceived in financial and IT auditing – and explore its implications for AI audits. The authors define continuous auditing of AI systems (CAAI) as a (nearly) real-time electronic support system for auditors that continuously and automatically audits an AI system to assess consistency with relevant norms and standards. In contrast with traditional audits, which tend to either be of either a discrete or a cyclical nature, CAAI changes the temporality of audits and affords real-time monitoring of current events. In their article, Minkkinen et al. demonstrate that CAAI is not only an understudied but also a promising methodological approach to identifying and managing the ethical and legal risks posed by AI systems operating with high degrees of autonomy or equipped with the capacity to 'learn' as they interact with dynamic environments over time.

In *The Self-Synchronisation of AI Ethical Principles*, Light and Panai (2022) take a step back to consider the principles against which AI systems are being audited. While many different sets of principles have been proposed by governments, NGOs and private sector actors, the authors argue that some degree of self-synchronisation is taking place. Further, they demonstrate how structured and independent audits can help facilitate this process of synchronisation of ethical principles. By promoting procedural transparency, regularity and verifiability, Light and Panai argue that audits contribute to an 'infrastructure of trust' that connects technology providers, users and society. The authors illustrate this process through a detailed case study of the Independent Audit of AI Systems (IAAIS) procedure developed by ForHumanity, a non-profit organisation. In their view, the task of auditors is not to intervene to change or align different organisations' ethical values but to support a plurality of ethical approaches to keep the process of self-synchronisation going.

Finally, in *Auditing of AI in Railway Technology – a European Legal Approach*, Gesmann-Nuissl and Kunitz (2022) highlight the challenges AI systems pose in the railway sector and outline an auditing procedure designed to verify AI systems used in that context. The authors argue that the opacity of machine-learning-based AI systems constitutes a major challenge for demonstrating functional safety in line with sector-specific railway regulations. As a potential solution, a procedure is proposed whereby the safety and functionality of AI systems are not verified analytically but by means of extensive testing. With that approach, it is not the capabilities of AI systems that are being audited but rather the processes whereby they are designed and deployed. Such a procedure, Gesmann-Nuissl and Kunitz conclude, would be consistent with both the conformity assessments mandated by the EU AIA and existing industry standards for software development.

## 6 Concluding remarks

This article has provided an overview of previous work on AI auditing. From this review, three key points have emerged. First, contemporary attempts to audit AI systems have much to learn from how audits have historically been structured and conducted in areas like financial accounting, safety engineering and the social sciences. Second, academic researchers can fill an important role by studying the feasibility and effectiveness of different AI auditing procedures. Third, auditing is an inherently multidisciplinary undertaking, whereby different approaches to auditing complement and mutually reinforce each other.

The contributions to Digital Society's topical collection surveyed in Section 5 support the above conclusions in different ways. To start with, Minkkinen et al. (2022) provide a good example of translational research, whereby best practices for continuous audits in financial and IT auditing are transposed into the context of AI auditing. More translational research is needed to ground emerging AI auditing procedures in the rigorous methodologies and cumulative experiences of audits in other domains.

Further, in Section 3, I demonstrated that the contemporary calls for AI systems to be audited result from a confluence of top-down and bottom-up pressures. To recap, both policymakers and technology providers have an interest in promoting auditing as a promising AI governance mechanism. The question is not whether an AI system will be audited, but whether these audits will be rigorous enough to provide adequate insurance against the risks AI systems pose. The task of studying the effectiveness and feasibility of different AI auditing procedures is thus one for academic researchers. Here, Hasan et al. (2022), Felländer et al. (2022) and Vetter et al. (2023) all make important contributions by (i) documenting the methodological affordances and constraints of different AI auditing procedures and (ii) reflecting on the challenges auditors and industry practitioners face when attempting to design and implement AI audits in applied settings.

Finally, AI auditing is an inherently multidisciplinary undertaking, which different researchers approach in different ways. Amongst others, it is possible to distinguish between legal, ethical and technical approaches. Gesmann-Nuissl and Kunitz (2022) approach the challenges associated with the use of AI systems in the railway sector from a legal point of view; Light and Panai (2022) conduct an ethical analysis of the principles against which AI systems are being audited; and Minkkinen et al. (2022) focus on the technical aspects of how audits are conducted. Importantly, these approaches are not mutually exclusive but rather critically complementary. For example, legal compliance audits typically rely on technical methods to gather evidence about the properties and impact of AI systems. Similarly, technical robustness and legal compliance are often prerequisites for considering an AI system ethical.

The main takeaway from this review article is that how AI auditing procedures are designed and implemented matters greatly. To be feasible and effective, AI auditing procedures should (i) be structured and transparent, (ii) assess a clearly defined material scope according to an equally clearly defined normative baseline, (iii) incorporate elements of both (narrow) technology-oriented assessments of ADMS and (broad) process-oriented assessments of organisations that design and deploy ADMS, (iv) include continuous monitoring of ADMS, and (v) be conducted by independent third-party auditors. However, even when conducted in line with these best practices, auditing – as an AI governance mechanism – is subject to a wide range of conceptual, technical, economic and institutional limitations. While some of these limitations can be addressed by appropriate policy responses and future technological innovation, others are intrinsic. Policymakers, researchers, and auditors should therefore exercise caution and remain realistic about what AI auditing can be expected to achieve.

This is a preprint. A revised version has been published in the journal *Digital Society*.

This is a preprint. A revised version has been published in the journal *Digital Society*.

This is a preprint. A revised version has been published in the journal *Digital Society*.